\documentclass[bm,aps,showpacs,amsfonts,amssymb]{revtex4}

\usepackage{graphicx,bm}
\usepackage{natbib}

\usepackage{color}
\begin{document}

{~}

\title{
New cylindrical gravitational soliton waves and gravitational Faraday rotation }
\vspace{2cm}
\author{Shinya Tomizawa${}^\dagger$ and Takashi Mishima${}^\ddagger$}
\vspace{2cm}
\affiliation{${}^\dagger $ University Education Center, Ibaraki University, 2-1-1, Bunkyo, Mito, Ibaraki, 310-8512, Japan, \\
${}^\ddagger$ Laboratory of Physics, College of Science and Technology, Nihon University,
Narashinodai, Funabashi, Chiba 274-8501, Japan}

\begin{abstract} 
In terms of gravitational solitons, we study gravitational non-linear effects of gravitational solitary waves such as Faraday rotation.
Applying the Pomeransky's procedure for inverse scattering method, which has been recently used for constructing stationary black hole solutions in five dimensions to a cylindrical spacetime in four dimensions, we construct a new cylindrically symmetric soliton solution. 
This is the first example to be applied to the cylindrically symmetric case. 
In particular, we clarify the difference from the Tomimatsu's single soliton solution, which was constructed by the Belinsky-Zakharov's procedure.
\end{abstract}

\pacs{04.50.+h  04.70.Bw}
\date{\today}
\maketitle

\section{Introduction}
Gravitational solitons in relativity are found in the framework of the inverse scattering method as exact solutions of the Einstein equations with two Killing vector fields. 
These solutions describe self-reinforcing solitary wave (a wave packet or pulse) that maintains its shape propagating in various physically interesting background space-times.  The gravitational soliton as a gravitational wave  interact the background itself as its non-linear effect of gravitation. 
On the other hand, in recent years,  the gravitational solitons have been studied in the context of constructing black hole solutions in higher dimensions.
In particular, Pomeransky~\cite{Pom} have improved the inverse scattering method which Belinsky-Zakharov~\cite{Belinsky-Zakharov,soliton book} established so that one can construct higher dimensional regular black hole solutions. 
Surprisingly, this improved method has succeeded in generating some physically interesting exact solutions which describe a certain kind of stationary and axisymmetric black holes~\cite{review IIM,review ER}. 
The main aim of this paper is to construct a new cylindrically symmetric gravitational soliton which describes gravitational waves by using the Pomeransky's procedure rather than by the Belinsky-Zakharov's original procedure~\cite{Belinsky-Zakharov}.

\medskip
Einstein and Rosen~\cite{Einstein-Rosen, book exact solution} provided the first cylindrical gravitational wave with a $+$ mode only, where the vacuum Einstein equation reduces to a simple linear wave equation.
Piran {\it et al}.~\cite{Piran} numerically studied time evolution and non-linear interaction of cylindrical gravitational waves of both polarization modes ($+$ and $\times$ modes) and showed that if an outgoing (ingoing) wave is linearly polarized when an ingoing (outgoing) $\times$-mode wave is present, the polarization for the outgoing (ingoing) wave rotates via the non-linear interaction as it propagates. Today this effect is well known as {\it gravitational Faraday effect}.
Tomimatsu~\cite{tomimatsu} studied the gravitational Faraday rotation for cylindrical gravitational solitons by using the inverse scattering technique. 
The outgoing $+$ waves emitted from an axis convert to $\times$ mode completely in the immediate interaction region and finally the outgoing waves contain both polarizations. 
For the single-soliton solution, the soliton disturbance exists only in the interior region of a future light cone and shock wave is present on the future light cone.
Further, to avoid the shock wave like structure on the light cone, he considered two-soliton solution with complex poles, which behaves like the single soliton field by a particular choice of the parameter. 
Moreover, the interaction of gravitational soliton waves with a cosmic string was also studied in~\cite{Economou,Xanthopoulos1,Xanthopoulos2}.

\medskip
In this paper, we study the non-linear interaction of new cylindrically symmetric soliton solutions which can be generated by the Pomeransky's procedure. 
As is well known, the most general metric for a cylindrically symmetric spacetime can be described by the Kompaneets-Jordan-Ehlers form~\cite{Komaneets-Jordan-Ehlers}.
Following Ref.~\cite{Piran}, in terms of the metric form, we calculate the wave amplitudes for the ingoing (outgoing) gravitational waves with the $+$ and $\times$ modes. 
According to Ref.~\cite{Piran,tomimatsu}, for such a cylindrically symmetric spacetime, the polarization angles of gravitational waves can be defined by the ratio of $\times$ wave amplitudes to $+$ wave ones. 
Using the useful definition and some convenient formulas in the reference~\cite{Piran,tomimatsu}, we will calculate time development of the polarization angles. 
We will find that for the single soliton solution, the $\times$ mode which is initially dominate near the axis of symmetry decreases with time and at a certain time, fully converts to the $+$ mode. 
After that, in turns, it begins to increase and finally again the $\times$ mode is completely dominant. 
Moreover, we will also show that the polarization vector of two independent modes changes as it propagating a background spacetime along an null geodesic (gravitational Faraday rotation). 

\medskip
In the next section, we present the metric in a general cylindrical spacetime 
and reduce the Einstein equation with the symmetry to two-dimensional equations.
In Sec.~\ref{sec:single}, using the improved inverse scattering method by the Pomeransky~\cite{Pom} rather than by the Belinski-Zakharov~\cite{Belinsky-Zakharov}, we will derive a single-soliton solution from Minkowski metric. In particular, we compute the amplitudes and polarization angles for ingoing and outgoing waves and clarify the difference from the Tomimatsu solution~\cite{tomimatsu} generated by the latter procedure. As discussed by Ref.~\cite{tomimatsu}, such a single soliton solution with real poles has as singular behavior on the light cone, 
which means that outgoing shock waves initially propagate at the speed of light velocity. 
Sec.~\ref{sec:discuss} is devoted to the summary of this paper and discussion on our results.

\section{Cylindrically symmetric spacetimes}\label{sec:cylinder} 
Throughout this paper, we assume that a four-dimensional spacetime admits two commuting Killing vector fields, an axisymmetric Killing vector $\partial/\partial \phi$ and a spatially translational Killing vector $\partial/\partial z$, where 
the polar angle coordinate $\phi$ and the coordinate $z$ have the ranges $0\le \phi<2\pi$ and $-\infty<z<\infty$, respectively.
The metric with cylindrical symmetry $U(1)\times {\bm R}$ is generally written in the Kompaneets-Jordan-Ehlers form: 
\begin{eqnarray}
ds^2=e^{2\psi}(dz+\omega d\phi)^2+\rho^2 e^{-2\psi}d\phi^2+e^{2(\gamma-\psi)}(d\rho^2-dt^2),
\end{eqnarray}
where the functions $\psi$, $\omega$ and $\gamma$ depend on the time coordinate $t$ and radial coordinate $\rho$ only. Following Ref.~\cite{tomimatsu}, we find that 
the function $\gamma$ is determined by
\begin{eqnarray}
&&\gamma_{,t}=\frac{\rho}{8}\left(A_+^2+B_+^2+A_\times^2+B_\times^2\right),\\
&&\gamma_{,\rho}=\frac{\rho}{8}\left(A_+^2-B_+^2+A_\times^2-B_\times^2\right),
\end{eqnarray}
where
\begin{eqnarray}
&&A_+=2\psi_{,v},\label{eq:Ap}\\
&&B_+=2\psi_{,u},\label{eq:Bp}\\
&&A_\times=\frac{e^{2\psi}\omega_{,v}}{\rho},\label{eq:Ac}\\
&&B_\times=\frac{e^{2\psi}\omega_{,u}}{\rho}.\label{eq:Bc}
\end{eqnarray}
Here, the advanced ingoing and outgoing null coordinates $u$ and $v$ are defined by $u=(t-\rho)/2$ and $v=(t+\rho)/2$, respectively.
The indices $+$ and $\times$ denote the respective polarizations.
The ingoing and outgoing amplitudes are defined by, respectively, 
\begin{eqnarray}
A=\sqrt{A_+^2+A_\times^2},\quad B=\sqrt{B_+^2+B_\times^2},\label{eq:amplitues}
\end{eqnarray}
and the polarization angles $\theta_A$ and $\theta_B$ for the respective wave amplitudes are given by, 
\begin{eqnarray}
\tan2\theta_A=\frac{A_\times}{A_+},\quad\tan2\theta_B=\frac{B_\times}{B_+}.\label{eq:pola}
\end{eqnarray}

\section{Single soliton solutions}\label{sec:single}
Using the Pomeransky's procedure~\cite{Pom} for the inverse scattering method, we construct a new single solitonic solution.
Let us choose Minkowski metric as a seed, whose $2\times 2$ part is written as
\begin{eqnarray}
g_0=\left(1,\rho^2\right).
\end{eqnarray}
First, remove a soliton with a vector $m^{(1)}=(1,0)$ at $t=0$ from this seed metric and then we get the metric $g_0'=(\mu^2/\rho^2,\rho^2)$.
Next, add back a non-trivial soliton with $m^{(1)}=(1,a)$ at $t=0$ to $g_0'$ and then we can obtain the metric of a single soliton solution.

\medskip
This is how we can obtain the metric coefficients for the single soliton solution, which is given by
\begin{eqnarray}
e^{2\psi}&=&\frac{(w^2-1)^4\rho^2+a^2w^4}{(w^2-1)^4\rho^2+a^2w^2},\\
\omega&=&-\frac{a(w^2-1)^3\rho^2}{(w^2-1)^4\rho^2+a^2w^4}, \label{eq:sol}\\
e^{2\gamma}&=&
\frac{a^2 w^4+\left(-1+w^2\right)^4 \rho ^2}{\left(-1+w^2\right)^4 \rho ^2}.
\end{eqnarray}
where $\mu:=\sqrt{t^2-\rho^2}-t$ and $w:=-\mu/\rho$. 
From eqs.(\ref{eq:Ap})-(\ref{eq:Bc}), it is straightforward to compute the respective wave amplitudes $A_+,B_+,A_\times$ and $B_\times$:
\begin{eqnarray}
&&A_+=-\frac{2 a^2 (w-1) w^3 \left(a^2 w^3-\rho ^2 (w-2) \left(w^2-1\right)^4\right)}{\rho  (w+1) \left(a^2 w^2+\rho ^2 \left(w^2-1\right)^4\right) \left(a^2 w^4+\rho ^2 \left(w^2-1\right)^4\right)},\\
&&B_+=-\frac{2 a^2 w^3 (w+1) \left(\rho ^2 (w+2) \left(w^2-1\right)^4-a^2 w^3\right)}{\rho  (w-1) \left(a^2 w^2+\rho ^2 \left(w^2-1\right)^4\right) \left(a^2 w^4+\rho ^2 \left(w^2-1\right)^4\right)},\\
&&A_\times= -\frac{2 a (w-1)^3 w^2 (w+1) \left(a^2 w^2 (2 w-1)+\rho ^2 \left(w^2-1\right)^4\right)}{\left(a^2 w^2+\rho ^2 \left(w^2-1\right)^4\right) \left(a^2 w^4+\rho ^2 \left(w^2-1\right)^4\right)},\\
&&B_\times=-\frac{2 a (w-1) w^2 (w+1)^3 \left(a^2 w^2 (2 w+1)-\rho ^2 \left(w^2-1\right)^4\right)}{\left(a^2 w^2+\rho ^2 \left(w^2-1\right)^4\right) \left(a^2 w^4+\rho ^2 \left(w^2-1\right)^4\right)}.
\end{eqnarray}
Hence, from eqs.(\ref{eq:amplitues}), the respective wave amplitudes are written as
\begin{eqnarray}
&&A=\frac{2 |a| |w-1| |w|^2}{\rho  |w+1| \sqrt{\left(a^2 w^4+\rho ^2 \left(w^2-1\right)^4\right)}},\\
&&B=\frac{2 |a| |w+1| |w|^2}{\rho  |w-1| \sqrt{\left(a^2 w^4+\rho ^2 \left(w^2-1\right)^4\right)}}.
\end{eqnarray}
From eqs.(\ref{eq:pola}), the polarization $\theta_A$ and $\theta_B$ angles are determined by
\begin{eqnarray}
&&\tan2\theta_A=\frac{\rho  \left(w^2-1\right)^2 \left(a^2 w^2 (2 w-1)+\rho ^2 \left(w^2-1\right)^4\right)}{a w \left(a^2 w^3-\rho ^2 (w-2) \left(w^2-1\right)^4\right)},\\
&&\tan2\theta_B=\frac{\rho  \left(w^2-1\right)^2 \left(a^2 w^2 (2 w+1)-\rho ^2 \left(w^2-1\right)^4\right)}{a w \left(\rho ^2 (w+2) \left(w^2-1\right)^4-a^2 w^3\right)}.
\end{eqnarray}

\section{Analysis of the single soliton solution}
Now let us investigate the new single soliton solution in detail.
It is useful to know how the restrictive wave components $A_+$, $B_+$, $A_\times$ and $B_\times$  and the polarization angles $\theta_A$, $\theta_B$ behaves near the boundaries of the spacetime, (1) the axis of symmetry $\rho=0$, (2) the light cone $t=\rho$, (3) the timelike infinity $t\to\infty$, and (4) null infinity $v\to\infty$.

\subsubsection{On the axis $\rho=0$}
\begin{figure}[htbp]
 \begin{center}
  \includegraphics[width=60mm]{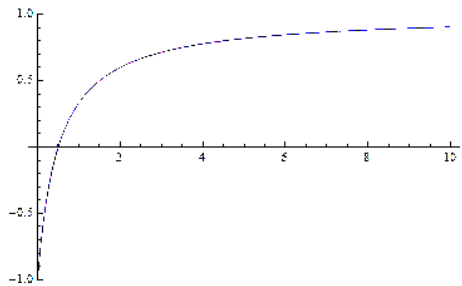}
 \end{center}
 \caption{The time dependence of $\tan\theta_A(=-\tan\theta_B)$ on the axis for ingoing (outgoing) waves, where we set $a=1$.}
 \label{pol}
\end{figure}

Near the axis of symmetry $\rho=0$, the restrictive amplitudes of ingoing and outgoing waves become
\begin{eqnarray}
A= B=\frac{|a|}{2t^2},
\end{eqnarray}
and the restrictive polarization angles take the values
\begin{eqnarray}
&&\tan\theta_A= -\tan\theta_B=\frac{a}{|a|}\frac{2t-|a|}{2t+|a|}.
\end{eqnarray}
 As shown in FIG.~\ref{pol}, 
 the polarization angle $\theta_A\ (=-\theta_B)$ of ingoing (outgoing) waves propagating on the axis depend on time $t$. 
 Initially, at $t=0$, the $+$ mode is absent and the pure $\times$ mode only are present. 
After that, the $+$ mode waves come to exist. 
In particular, when $t=|a|/2$, the $\times$ mode completely vanish and the $+$ mode only is present.
Further, after that, the $+$ mode converts to the $\times$ mode and at $t\to\infty$, approaches to the $\times$ mode completely. 
Note also that for the solution, in~\cite{tomimatsu}, on the axis, the $\times$ mode are always absent. 

\medskip
We see that the $C$-energy density is proportional to
\begin{eqnarray}
\gamma_{,\rho}&=&\frac{a^2 w^4 \left(1+6 w^2+w^4\right)}{\rho\left(-1+w^2\right)^2 \left(a^2 w^4+\left(-1+w^2\right)^4 \rho ^2\right)}\\
               &\simeq&\frac{a^2}{16t^4}\rho. \label{eq:cenergy}
\end{eqnarray}
Therefore, in contract to the single soliton in Ref.~\cite{tomimatsu}, the $C$-energy does not diverge on the axis. it is regular on the axis (for the solution in~\cite{tomimatsu}, it diverges). 

\medskip
Near the axis, the metric behaves as
\begin{eqnarray}
ds^2\simeq \left(1+\frac{a^2}{4t^2}\right)^{-1}\left(dz+ad\phi\right)^2+\rho^2\left(1+\frac{a^2}{4t^2}\right)d\phi^2+\left(1+\frac{a^2}{4t^2}\right)(-dt^2+d\rho^2).
\end{eqnarray}

Let us introduce a new coordinate $\tilde z:=z+a\phi$. Then the asymptotic behavior of the metric can be written as
\begin{eqnarray}
ds^2\simeq \left(1+\frac{a^2}{4t^2}\right)^{-1}d\tilde z^2+\rho^2\left(1+\frac{a^2}{4t^2}\right)d\phi^2+\left(1+\frac{a^2}{4t^2}\right)(-dt^2+d\rho^2).
\end{eqnarray}
Therefore, under this choice of the periodicity $\Delta\phi=2\pi$, there is no deficit angle for the solution.

 \medskip
\subsubsection{On light cone ($t=\rho$)}
Since the amplitudes of the ingoing waves $A_+$ and $A_\times$ vanish on the light cone $w=1\ (t= \rho)$ (see eqs.(\ref{eq:Ap}) and (\ref{eq:Ac})), 
we see that no ingoing wave crosses it. 
On the other hand, for the outgoing waves, $B_\times$ vanishes on it but
$B_+$ diverges (see eqs.(\ref{eq:Bp}) and (\ref{eq:Bc})), therefore, the $+$ mode with infinite wave amplitude only exists there. 
Hence, from this fact and Eq.~(\ref{eq:cenergy}), we can read off that the $C$-energy density does also diverges on it.
This singular behavior on the light cone is similar to that in~\cite{tomimatsu}, where the largest portion of the disturbance for the outgoing wave with both $+$ and $\times$ modes lies on the light cone. 
The largest portion of the outgoing gravitational radiation can be regarded as a gravitational wave pulse propagating at the speed of light from the axis $(t,\rho)=(0,0)$ to the null infinity. 
As will be seen later, it should be noted that the gravitational shock wave starting from the axis interact with the background spacetime and generate the outgoing $\times$ mode wave and ingoing waves.

\medskip

\begin{figure}[h]
 \begin{center}
  \includegraphics[width=60mm]{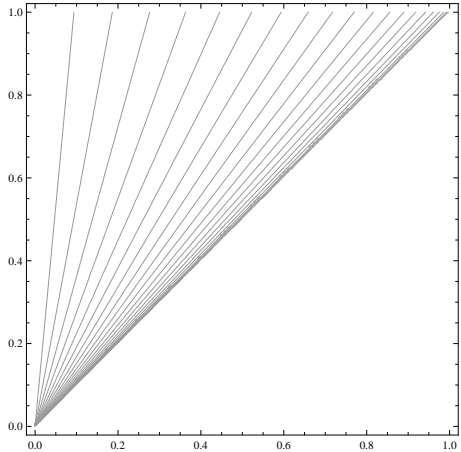}
 \end{center}
 \caption{$w={\rm constant}$ curves in the $(t,\rho)$-plane: each $w={\rm constant}$ curve denotes a straight line. The axis $\rho=0$  and the light cone $t=\rho\ (u=0)$ correspond to the lines $w=0$ and $w=1$, respectively.}
 \label{fig:w}
\end{figure}

\subsubsection{Timelike infinity $t\to\infty$}
Let us consider the limit $t\to\infty$ with keeping the radial coordinate $\rho$ constant. 
It is easy to confirm that the spacetime at the timelike infinity  asymptotically behaves as Minkowski spacetime, actually, at $t\to\infty$ the metric behaves as
\begin{eqnarray}
ds^2\simeq 
\left(1-\frac{a^2}{4t^2}\right)\left\{dz+a\left(1+\frac{\rho^2}{4t^2}\right)d\phi\right\}^2+\rho^2\left(1+\frac{a}{4t^2}\right)d\phi^2+\left(\rho^2\sin^2\theta +\frac{1}{t^2}\right)d\phi^2+\left(1+\frac{a^2}{4t^2}\right)(-dt^2+d\rho^2).
\end{eqnarray}

Hence, this implies both the amplitudes of the ingoing and outgoing gravitational waves gradually decay and finally vanish completely :
\begin{eqnarray}
&&A\simeq \frac{a^2}{2t^2}+{\cal O}(t^{-3}),\\
&&B\simeq \frac{a^2}{2t^2}+{\cal O}(t^{-3}).
\end{eqnarray}
We also find that $\tan\theta_A$ and $\tan\theta_B$ for the ingoing and outgoing waves behave as, restrictively, 
\begin{eqnarray}
\tan\theta_A\simeq -\tan\theta_B\simeq 1,
\end{eqnarray}
which means that for fixed the radial coordinate $\rho$, the $\times$ mode for  the restrictive  waves  becomes dominate as time passes.

\subsubsection{Null infinity $v\to \infty$ with $u=u_0>0$}
We can see that at the null infinity $v\to \infty$ with $u=u_0>0$, the spacetime approaches to Minkowski spacetime. 
In this limit, the metric at $v\to\infty$ asymptotically behaves as
\begin{eqnarray}
ds^2\simeq \left(dz+\frac{a}{4u_0^{1/2}b^2 }v^{1/2}d\phi\right)^2+\rho^2d\phi^2+b^2(-dt^2+d\rho^2),
\end{eqnarray}
where $b:=\sqrt{1+(a/16u_0)^2}$. This asymptotic metric form seems to be singular at the infinity because of existence of the additional parameter $a$. 
To see that actually, the metric is asymptotically flat, let us introduce new coordinates $(\tilde t,x,y)$ defined by
\begin{eqnarray}
&&\tilde t=bt,\\
&&x=b\rho\cos b\phi,\\
&&y=b\rho\sin b\phi,
\end{eqnarray}
and then we can see that at $v\to\infty$ the metric behaves as
\begin{eqnarray}
ds^2\simeq -d\tilde t^2+dx^2+dy^2+dz^2+ {\cal O}(v^{-1/2}).
\end{eqnarray}
Therefore, at the null infinity, both the ingoing and going waves asymptotically vanishes.

\medskip
Next, see how the ingoing and outgoing waves behave along null rays $u=u_0(>0)$ by changing a value of $u_0$. 
Here, without loss of generality, we can assume $a>0$ because under the transformation $a\to-a$ and $\phi\to-\phi$, the single soliton solution is invariant. 
Note that at the null infinity $v\to\infty$, the restrictive wave amplitudes behave as
\begin{eqnarray}
&&A\simeq \frac{2|a|u_0}{(256u_0^2+a^2)^{1/2}v^{3/2}},\\
&&B\simeq \frac{2|a|}{\{(256u_0^2+a^2)u_0\}^{3/2}v^{1/2}},
\end{eqnarray}
and the restrictive polarization angles $\theta_A$ and $\theta_B$ behave as
\begin{eqnarray}
&&\tan\theta_A \simeq \frac{-a+(256u_0^2+a^2)^{1/2}}{16u_0},\\
&&\tan\theta_B \simeq 
\frac{16u_0(256u_0^2-3a^2)}{a(a^2-768u_0^2)+\epsilon(256u_0^2+a^2)^{3/2}},
\end{eqnarray}
where $\epsilon$ takes $1$ when $u_0<|a|/(16\sqrt{3})$ and $-1$ when $u_0>|a|/(16\sqrt{3})$.

\subsubsection{Gravitational Faraday effect}
It is physically interesting to see how the polarization angle $\theta_B$  of the outgoing wave propagating from the axis $\rho=0$ changes along some null rays $u=u_0$ by changing a value of $u_0$. 
This effect is well known as gravitational Faraday rotation.
 Note that at $v\to\infty$, $\tan\theta_A$ is always positive and the signature of $\tan\theta_B$ depends on the value of $u_0$.  
Also note that on the axis $\rho=0$,  $\tan\theta_A=-\tan\theta_B$ is negative when $0<u_0<a/4$, positive when $u_0>a/4$, and zero when $u_0=a/4$.\\

(i) For $u_0<|a|/16\sqrt{3}$ ($t<|a|/8\sqrt{3}$ on the axis $\rho=0$), both $+$ and $\times$ modes are present on the axis.
The ratio of the $\times$ mode wave to the $+$ mode wave is decreasing along a null ray $u=u_0$ and the outgoing wave converts to $B_+$ completely.
After that, a part of the pure $+$ mode converts to the $\times$ mode and the ratio is increasing. 

\medskip
(ii) For $|a|/16\sqrt{3}<u_0<\sqrt{3}|a|/16$ ($\sqrt{3}|a|/8<t<\sqrt{3}|a|/8$ on the axis $\rho=0$), both $+$ and $\times$ modes are present on the axis. 
The ratio of the $\times$ mode wave to the $+$ mode wave is decreasing along a null ray $u=u_0$ but does not vanish anywhere. 

\medskip
(iii) For $u_0=\sqrt{3}|a|/16$ ($t=\sqrt{3}|a|/8$ on the axis $\rho=0$), both $+$ and $\times$ modes are present on the axis. 
The ratio of the $\times$ mode wave to the $+$ mode wave is decreasing along a null ray $u=u_0$ and vanishes at infinity $v\to\infty$. 
Therefore, at null infinity, the $+$ mode wave only is present.

\medskip
(iv) For $\sqrt{3}|a|/16<u_0<|a|/4$ ($\sqrt{3}|a|/8<u_0<|a|/2$ on the axis $\rho=0$), the behavior of outgoing wave is similar to the case of (i).

\medskip
(v) For $u_0=|a|/4$ ($t=|a|/2$ on the axis $\rho=0$), the $\times$ mode wave vanishes on the axis.
The pure $+$ mode wave emitted from the axis partially converts to the $\times$ mode and then the ratio is increasing along a null ray.

\medskip
(vi) For $|a|/4<u_0$ ($|a|/4<t$ on the axis $\rho=0$), both $+$ and $\times$ modes are present on the axis. 
The ratio of the $\times$ mode wave to the $+$ mode wave is increasing along a null ray $u=u_0$.

\medskip
Next, see how the polarization angle $\theta_A$  of the ingoing wave from the axis $\rho=0$ to the infinity $v\to\infty$ changes along some null rays $u=u_0$ by a value of $u_0$. 

\medskip
(i) For $u_0<|a|/4$ ($t<|a|/2$ on the axis $\rho=0$), both $+$ and $\times$ modes are present on the axis. 
The ratio of the $\times$ mode wave to the $+$ mode wave is decreasing along a null ray $u=u_0$ and the outgoing wave converts to $B_+$ completely.
Then, a part of the pure $+$ mode converts to $\times$ and the ratio is increasing. 

\medskip
(ii) For $u_0=|a|/4$ ($t=|a|/2$ on the axis $\rho=0$), the $\times$ mode wave vanishes on the axis. 
The pure $+$ mode wave emitted from the axis partially converts to the $\times$ mode and then the ratio is increasing along a null ray.

\medskip
(iii) For $u_0>|a|/4$ ($t>|a|/2$ on the axis $\rho=0$), both $+$ and $\times$ modes are present on the axis. 
The ratio of the $\times$ mode wave to the $+$ mode wave is increasing along a null ray $u=u_0$ at large $v\gg |a|$ and does not vanish and does not reach $1$ anywhere.

\section{Discussion}\label{sec:discuss}

In this paper, applying the Pomeransky's procedure for the inverse scattering method to a cylindrically symmetric spacetime, we have obtained the gravitational soliton as an exact solution 
to vacuum Einstein equations with cylindrical symmetry. 
We would like to emphasize that this work is the first example to be generated by the Pomeransky's procedure in such a cylindrical context. 
In general, such a single-soliton describes gravitational wave pulse with time-depending polarization angles propagating through a cylindrically symmetric spacetime.
In terms of the gravitational soliton, we have studied the effect of the gravitational Faraday rotation of gravitational waves. 
In particular, we have compared our single soliton solutions with the Tomimatsu's single-soliton~\cite{tomimatsu} which was constructed by the use of the original Belinsky-Zakharov's procedure. 

\medskip
Here, we would like to point out that there are essential differences between the two solutions: 

\medskip
(i) For our solution, the $C$-energy diverges on the light cone $u=0$ as for the Tomimatsu's solution. 
The outgoing wave with the largest portion of the disturbance initially propagates at light velocity and hence we can physically interpret 
it as a gravitational wave pulse.  
It should be noted that 
for the Tomimatsu's solution both the $+$ and $\times$ mod have infinite wave amplitude with the polarization fixed, while for our solution the outgoing wave with the $+$ mode only has infinite amplitude and one with the $\times$ mode vanishes there. 

\medskip
(ii) For our solutions, the $C$-energy does not diverge on an axis of symmetry, which is contrast to the soliton solutions in Ref.~\cite{tomimatsu}. 
This difference comes from the fact that for our solutions, the seed added back trivial soliton(s),i.e., Minkowski has no singular behavior on the axis. 
It is expected that the Pomeransky's procedure does not change this structure by the solitonic transformations except the point where the soliton is added. 
For the Tomimatsu's solution, the singular source on the axis continues to absorb and emit gravitational waves 
constantly (therefore, amplitudes does not decay near the axis forever).  
On the other hand, for our solution, because of absence of any sources on the axis, shock wave is emitted initially from the origin of the spacetime
and continues to scatter waves backward and make a tail, which is gradually damped and finally vanishes. 

\medskip
(iii) For our solutions, the polarization angles $\theta_A$ and $\theta_B$ of gravitational waves on the axis have time-dependence. 
This behavior is contrast to the soliton solutions in Ref.~\cite{tomimatsu}, where the $+$ mode amplitudes are dominant on the axis. 
At $t=0$, the pure $\times$ mode amplitudes only are present on the axis.
As time goes on, the $+$ mode gradually increases, and at some instance the $\times$ mode completely vanishes and only the $+$ mode is left. 
Then the $+$ mode is decreasing in turn and finally vanishes, while the $\times$ mode only remains. 
It should be noticed that the time-dependence of the polarization of gravitational waves on the axis is determined by the behavior of the polarization of the ingoing waves near the axis, because the outgoing waves are derived just by reflection of the ingoing waves due to the regularity of the axis. 

\medskip
(iv) At $t\to\infty$ (with $\rho$ constant), for our solutions,
 the spacetime asymptotically approaches to Minkowski, and simultaneously both ingoing and outgoing gravitational waves fade into the background spacetime. 
We also find that the $\times$ mode for the ingoing and outgoing waves becomes dominantin the future. On the other hand, for Tomimatsu's solutions, the spacetime is not asymptotically Minkowski at timelike infinity, and the waves does not vanish anywhere due to the existence of the singular source on the axis as mentioned above in  (ii). 
The polarizations approach to $0$ for both ingoing and outgoing waves, and therefore the $+$ modes gradually become dominant in the future. 

\medskip

Among the properties mentioned above, one of the most peculiar phenomena may be the $\times$ mode dominance of the wave tail.
As is well-known, all of the Einstein-Rosen wave pulses, which bear the $+$ mode only, have the tails which approach to the timelike infinity in the same way \cite{Rosen}.
The asymptotic forms of the ingoing and outgoing $+$ mode amplitudes behave as $1/t^2$.
On the other hand, for the wave pulse treated here, the $\times$ mode tails for the ingoing and outgoing waves have the same asymptotic form as the Einstein-Rosen wave pulses, but the $+$ mode tails decay more swiftly than the case of Einstein-Rosen wave pulses.
It seems to be plausible that the $\times$ mode plays the role of $+$ mode of the Einstein-Rosen wave pulses.
Actually we may support this statement with the aid of the following basic equations of $+$ mode amplitudes $( A_+,\, B_+ )$ and $\times$ mode amplitudes $( A_\times,\,B_\times )$ given in \cite{Piran}:
$$
{A_{+}}_{,u}={B_{+}}_{,v}=\frac{A_{+}-B_{+}}{2\rho}+A_{\times}B_{\times}\,,
$$
$$
{A_{\times}}_{,u}=\frac{A_{\times}+B_{\times}}{2\rho}-A_{+}B_{\times},\ \ {B_{\times}}_{,v}=-\frac{A_{\times}+B_{\times}}{2\rho}-B_{+}A_{\times}\,.
$$
For the right hand side of each equation above, the first term is the `linear' term which exists in the original equations of the Einstein-Rosen waves and the second one is `non-linear' term which gives new effects absent in Einstein-Rosen waves like the gravitational Faraday rotation.
Substituting the exact form of the new solution into the right hand side of the above equations,
we can easily evaluate which term becomes dominant between the linear and non-linear terms as time goes to infinity.
From this simple evaluation, we know that for the $\times$ mode the first term dominates second one so that the linear term controls the behavior of the $\times$ mode amplitudes, while for the $+$ mode the second term keeps the size comparable to the first term so that some non-linear effect still has some influence on the behavior of the $+$ mode.
As a result, we may suggest that the $\times$ mode behaves like the case of the Einstein-Rosen waves, but the $+$ mode rapidly decreases owing to the non-linear effect originating from the second term of the right hand side of the equations.
To study this further, it is interesting to clarify to what extent this phenomenon occurs generally.
From this point, we must treat more general cases including other soliton solutions and non-soliton solutions.

\end{document}